\documentclass{article}
\usepackage{amsmath}
\usepackage{graphicx}
\usepackage{hyperref}
\usepackage{subcaption}
\usepackage{caption}
\usepackage[numbers,sort&compress]{natbib}
\usepackage[margin=1in]{geometry}

\title{Parallel Architecture of a Frequency Comb Qudit Quantum Processor}

\author{
Duncan L. MacFarlane, Hiva Shahoei, \\
Murphy V. Paul, R. Mason Tuller, \\
and Mitchell A. Thornton \\
The Darwin Deason Institute for Cybersecurity \\
and \\
Department of Electrical and Computer Engineering \\
The Bobby B. Lyle School of Engineering \\
Southern Methodist University \\
Dallas, Texas \\
Corresponding author: Duncan MacFarlane, \href{mailto:dmacfarlane@smu.edu}{dmacfarlane@smu.edu}
}
\date{}

\begin{document}

\maketitle

\begin{abstract}
Quantum optical frequency combs provide an intriguing approach to high-dimensional quantum states. Because of the need to move probabilities among different colors, the realization of gates appropriate to multicolor photons requires nonlinear or electro-optic mixing. This paper describes a novel architecture for such gates. The parallel arrangement of mixers by dimension allows graceful scaling beyond two-dimensional qubits. The parallelism of the implementation simplifies the programming of the gate for a particular operation. As an example, we demonstrate the design of a four-dimensional Chrestenson operator.   

Topic: Integrated Photonics; Multicolor Photons, Quantum Informatics
\end{abstract}

\section{Introduction}

When one articulates photonic quantum processing with polarization-encoded qubits, components such as wave plates, beam splitters, and polarizers are employed. Similarly, quantum circuits for spatially encoded photonic qudits use couplers, interferometers, ring resonators, and combiners. The transfer functions for all the components identified so far, and hence the circuits they enable, are linear and time-invariant (LTI). 

Working with multicolor photon states requires the movement of probabilities among the colors (wavelengths or frequencies) \citep{olislager2010, chen2014, roslund2014, kobayashi2016, kues2017, mahmudlu2023, clementi2023}. LTI components will not accomplish this task. Consequently, nonlinear optics effects, or more commonly, mixing, become necessary operations in the construction of quantum gates for multicolored photon qubits and qudits. In many works to date, phase modulators enable the mixing of photon colors, and linear time-invariant filters adjust the amplitude and phases of the components produced by the RF electro-optic modulators \citep{harris2008, sensarn2009, lukens2017, kues2017, lu2018, imany2018, lu2019, kues2019, lu2023}. 

The purpose of this paper is to present an alternative to the predominantly serial architecture of EOMs and LTI filters favored to date to realize quantum logic gates for processing multicolor photon states. The architecture described herein scales gracefully to states of higher dimension, providing a flexible platform for advanced quantum information processing. The architecture described herein scales gracefully to states of higher dimension, has a desirable geometry and footprint when implemented as a quantum photonic integrated circuit (QPIC) cell, and has a preferable overall loss characteristic in comparison to the ``quantum Fourier processor'' (QFP) architecture as described in \citep{lukens2017,lu2018,kues2019}. 

\section{A Parallel Architecture}

\begin{figure}
    \centering
    \includegraphics[width=0.75\linewidth]{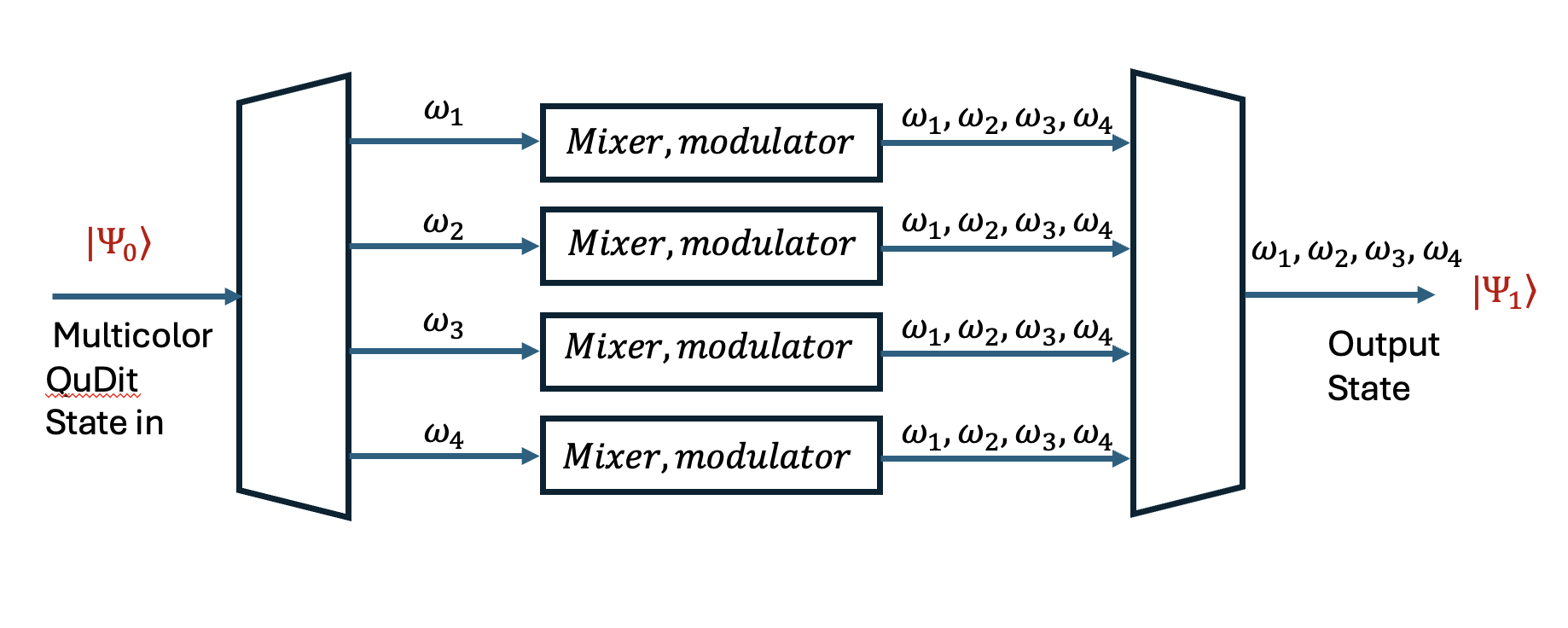}
    \caption{Qudit quantum frequency processor. A four color qudit is input to the gate. A demux separates the colors and the demultiplexed components are mixed in a parallel arrangement of  MZMs. The mixed components exit the gate through a mux. While a four-dimensional state is illustrated, more mixers may be added in parallel to scale gracefully to higher dimension.}
    \label{fig:QQFP}
\end{figure}

In this work, two modifications of past circuits are employed. LTI filters and Electro-optic modulators are still deployed; however, more complicated modulators – Mach-Zehnder Modulators (MZMs) and nested MZMs – are brought to bear. Secondly, these integrated mixers and filters are arranged in a parallel architecture. This is shown in Figure~\ref{fig:QQFP}.

An input multicolor photon state enters the quantum gate from the left. In a manner similar to the grating in the programmable filter in \citep{lukens2017,lu2018,kues2019}, a demultiplexer (DeMUX) parallelizes the constituent frequencies (color components). Since most multicolor photon states entail combs of frequencies defined by some free spectral range (FSR) of the generation component, the DeMUX may be realized with LTI filters, ring resonators, or cascades of Mach-Zehnder Interferometers (MZIs). The former is an infinite input response (IIR) filter, and this has ramifications for the effects of its group delay. The latter, finite impulse response (FIR) filter may require a multiplicity of stages to achieve a desired frequency discrimination, with concomitant overheads of control and parasitic losses. In either case, knowledge of the comb frequency spacing greatly simplifies this DeMUX. 

Each frequency component is then modulated by a specified RF signal, $m(t)$. For a $K$-order quantum state, each RF modulating signal typically comprises $(K-1)$ cosine components, each with designated amplitudes and phases. Since the MZMs and nested MZMs have multiple arms, each electro-optic segment may be driven by its own appropriate RF cosine Fourier series. Thus, the amplitude and phases provide degrees of freedom that may be designed to achieve the required transfer matrix of the gate at issue.

A final weight and phase may be assigned to each frequency of the multicolor photon before multiplexing together to provide the gate's output state \(|\Psi\rangle \).

The use of MZMs integrates mixing and filtering functionalities, since an MZM is a time-varying FIR filter. Thus, the architecture combines the mixing and filtering operations that are distinct in the serial architectures of \citep{lukens2017,lu2018,kues2019}.

\section{The realization of a Chrestenson $(\bf{C}_4)$ gate}

As an illustration of the architecture, and its applicability to states with dimension higher than 2, we consider the realization of the Chrestenson (\(\bf{C}_4\)) gate. The \(\bf{C}_4\) gate is the 4-dimensional analog of the Hadamard gate, a beam splitter for two-dimensional qubits. A unitary transfer matrix for the \(\bf{C}_4\) gate is:

\begin{equation}
\mathbf{C}_{4}=\frac{1}{\sqrt{4}}\left[\begin{array}{rrrr}
1 & 1 & 1 & 1 \\
1 & i & -1 & -i \\
1 & -1 & 1 & -1 \\
1 & -i & -1 & i
\end{array}\right]
\label{eq:c4-matrix}
\end{equation}

Functionally, each basis state component incident to the $\mathbf{C}_4$ gate is mixed to contribute equally to each output basis component in terms of their probability amplitudes. The induced phase shifts of this mixing ensure that the transfer matrix properly realizes the $\mathbf{C}_4$ operator as a unitary so that energy is conserved. Because $\mathbf{C}_4$ is a normalized four-dimensional discrete Fourier transform (DFT) matrix, it can be easily expanded into a higher $D$-dimensional form, $\mathbf{C}_D$, through use of the well-known DFT matrix properties as shown in Eqs. \ref{eq:roots-of-unity} and \ref{eq:general-DFT}.

\begin{equation} \label{eq:roots-of-unity}
w_m^k=e^{i\frac{2{\pi}km}{D}}
\end{equation}

\begin{equation} \label{eq:general-DFT}
\mathbf{C}_D=\frac{1}{\sqrt{D}}
\left[\begin{array}{rrrr}
w_0^0 & w_1^0 & \cdots & w_{D-1}^0 \\
w_0^1 & w_1^1 & \cdots & w_{D-1}^1 \\
\vdots & \vdots & \ddots & \vdots \\
w_0^{D-1} & w_1^{D-1} & \cdots & w_{D-1}^{D-1}
\end{array} \right]
\end{equation}

For the case of path-encoded qudits, a LTI 4-port coupler can be used to implement the \(\mathbf{C}_4\) operator \cite{macfarlane2004}\cite{macfarlane2011}\cite{smith2018}. However, the basis state components for a multicolor photon's quantum state are the colors, or energies/frequencies, of the photon. Hence, mixing is achieved by driving the MZMs with the appropriate electrical RF waveforms. In this example, we drive each branch modulator within the MZM with an RF signal that is conveniently represented as a three-term Fourier series,
\begin{equation}
\label{eq:fourier-series}
m(t) = a_1 \cos(\omega_{\text{m1}} t + \phi_{\text{1}}) + a_2 \cos(\omega_{\text{m2}} t + \phi_{\text{2}}) + a_3 \cos(\omega_{\text{m3}} t + \phi_{\text{3}}),
\end{equation}
that causes the MZM to spread each basis state amplitude component from one color to the others. In this work, we define our modulation signals as \( \omega_{\text{m1}} = \delta,\) \(\omega_{\text{m2}} = 2\delta,\) and \(\omega_{\text{m3}} = 3\delta \), where \( \delta \) is the RF frequency of the Free Spectral Range (FSR) defining the pure state structure of the frequency-encoded qudit of interest. Another commonly used component that generates photons with superimposed frequencies (not shown) is in the form of a frequency comb with a FSR of $\delta$.

\begin{figure}[htbp]
    \centering
    \begin{subfigure}[t]{0.45\textwidth}
        \centering
        \includegraphics[width=\linewidth]{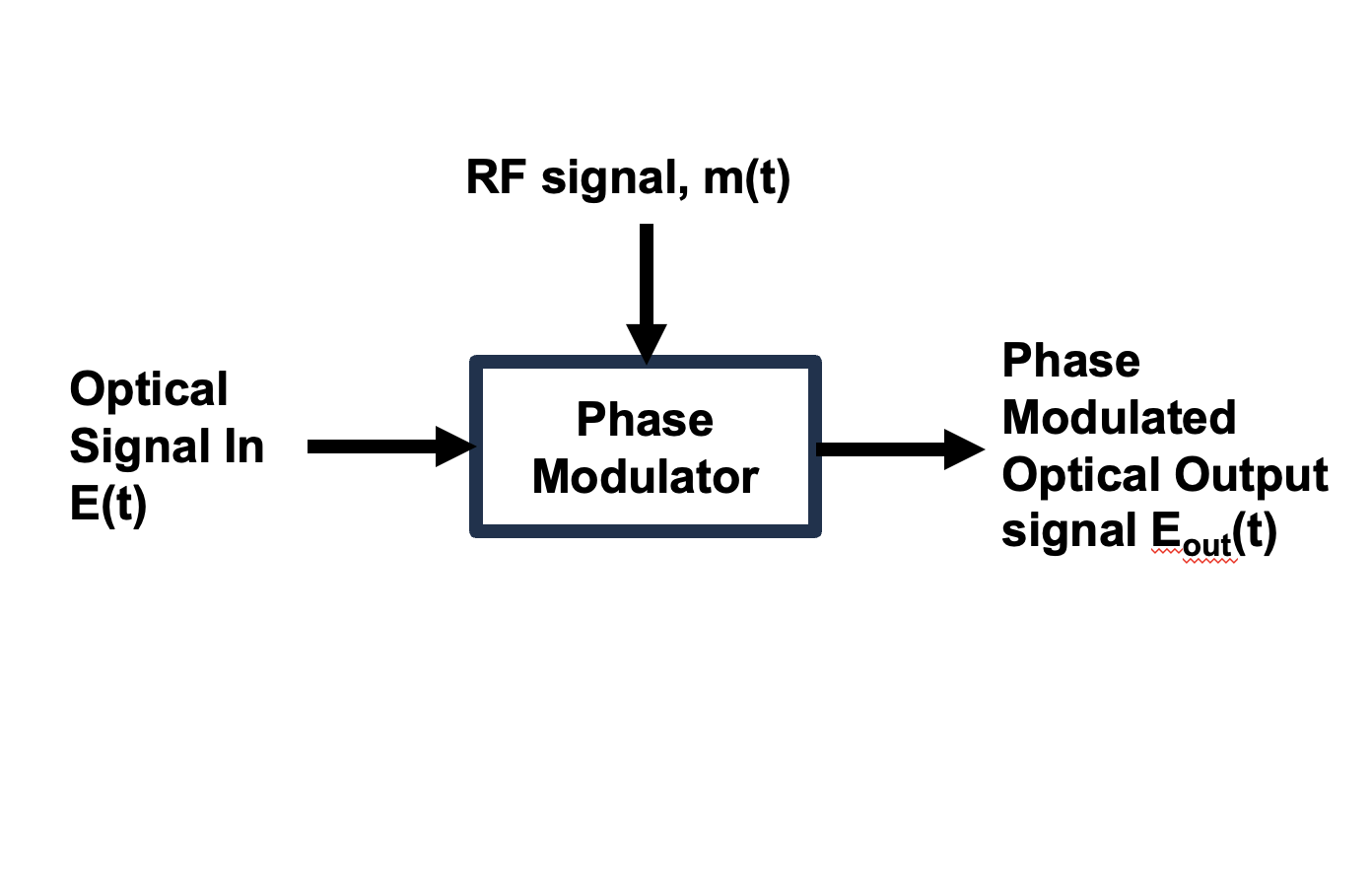}
        \caption{A single arm electro-optic modulator. An RF signal modulates the index of refraction of the material carrying the wave packet. This leads to a phase modulaton with an output that is a Bessel series expansion of probabilities across harmonics.}
        \label{fig:diagram1}
    \end{subfigure}
    \hfill
    \begin{subfigure}[t]{0.45\textwidth}
        \centering
        \includegraphics[width=\linewidth]{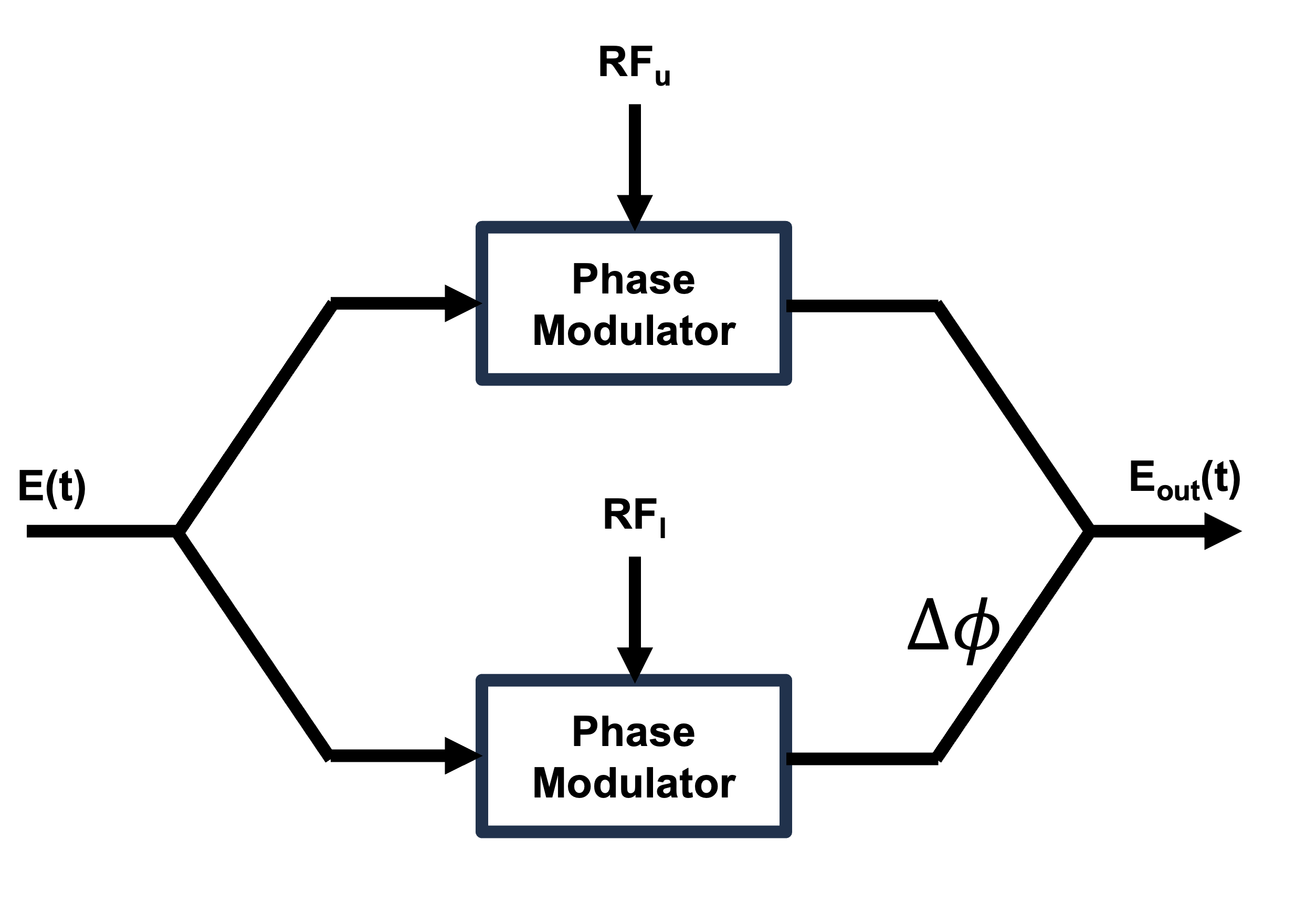}
        \caption{A mach-Zehnder modulator (MZM) can be viewed as two phase modulators in parallel.}
        \label{fig:diagram2}
    \end{subfigure}
    
    \vspace{1em}
    
    \begin{subfigure}[t]{0.9\textwidth}
        \centering
        \includegraphics[width=0.9\linewidth]{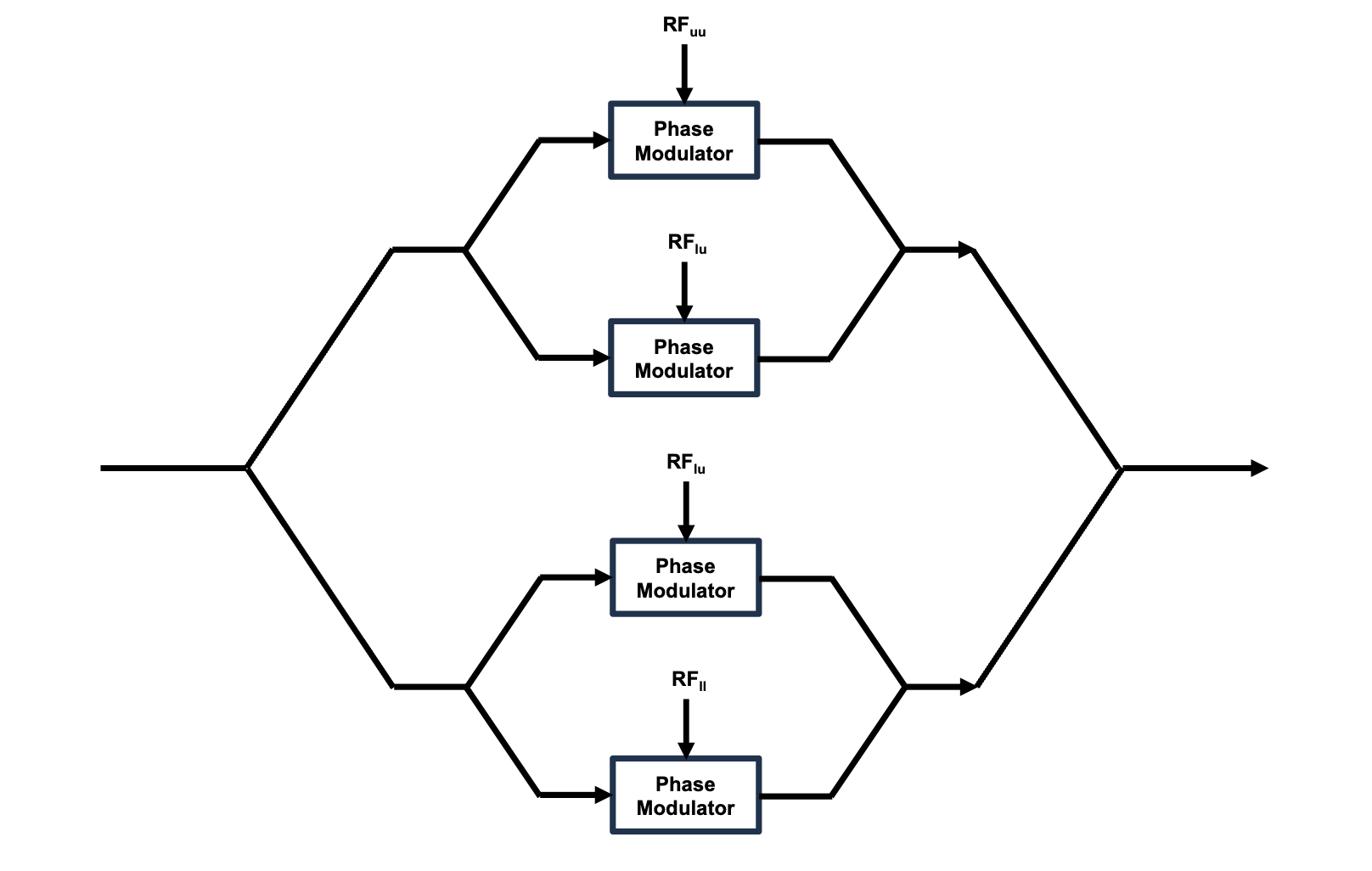}
        \caption{A nested MZM providing more degrees of freedom for mixing and filtering of a multicolor photon state}
        \label{fig:diagram3}
    \end{subfigure}
    \caption{Electro-optic modulators considered in this work.}
    \label{fig:all_diagrams}
\end{figure}

Figure~\ref{fig:diagram1} depicts a symbol representing an electro-optical phase modulator (EOM) typically used in multicolor photon quantum processing. When a three-tone modulation signal, $m(t)$, of the form of Eq.~\eqref{eq:fourier-series} is applied to the EOM, an optical input denoted as $E(t)$ produces a Bessel series output $E_{out}(t)$ of the form, 
\begin{equation}
\label{eq:output_inf_sum}
    E_{out}(t) = \sum_{n=-\infty}^{\infty}\sum_{m=-\infty}^{\infty}\sum_{k=-\infty}^{\infty}J_n(\beta_1)J_m(\beta_2)J_k(\beta_3)\cos(\omega_ct+n\omega_\text{m1}t+m\omega_\text{m2}t+k\omega_\text{m3}t+n\phi_1+m\phi_2+k\phi_3)
\end{equation}

An MZM comprises two such phase modulators, one in an upper arm and the other in a lower arm. This arrangement is shown in Figure~\ref{fig:diagram2}. Each arm produces an output that is of the form of Eq.~\eqref{eq:output_inf_sum} and these two expressions are added together with an appropriate phase shift, $\Delta\phi$. The use of an MZM over a single arm phase modulator provides more degrees of freedom that can be used to enhance the performance of the gate. 

If one desires more such control, then additional MZM arms may be deployed, for example, the nested MZM structure is widely used in modern coherent photonic communication systems. A nested MZM can form any desired Quadrature Amplitude Modulated (QAM) state \cite{winzer2006}. A nested MZM is shown in Figure~\ref{fig:diagram3}. This setup consists of two MZIs arranged inside a larger one. The nested design allows for more precise control of phase shifts and interference patterns, enabling enhanced modulation and filtering capabilities.

In the next two sections, we approach this example using two complementary approaches. First, a genetic annealing algorithm provides a numerical (local) optimization. Secondly, an analytic approximation is presented.

\section{Numerical Optimization}

\subsection{Genetic Algorithm}
Genetic algorithms (GA) are widely used statistical heuristic optimization methods inspired by biological processes such as natural selection, mutation and genetic crossover that allow for new populations of problem solutions to evolve as introduced in \cite{Hol92-genetic-algorithms}. 
The cost function contains two terms that simultaneously focus on the desirable in-band spectral lines and the undesirable out-of-band spectral lines or ``leakage.''  For the in-band components, we employ Eq.~\eqref{eq:output_inf_sum} to calculate the output of the nested MZM and use a GA to minimize the Frobenius norm among the ideal $\mathbf{C}_4$ transfer matrix and that obtained from the circuit.  This portion of the cost function, $c_{in}$ is applied to the in-band spectral components and is shown in Eq.~\ref{eq:in-ga-cost}.

\begin{equation}  \label{eq:in-ga-cost}
c_{in} = |\mathbf{C}_4 - \mathbf{T}_{circ}|^2
\end{equation}

The out-of-band cost, $c_{out}$, is also minimized simultaneously with the overall GA cost function, $c_{ga}=c{in} + c_{out}$ being minimized.  The leakage is computed as the energy or squared amplitude for the out-of-band spectral lines and the out-of-band cost, $c_{out}$ is just the sum of all the leakages.  We let that GA evolve 1,000 generations to obtain the parameter values that produce results closest to the desired amplitude and phase of the in-band frequency sidebands, while minimizing probabilities in the other sidebands due to the $c_{out}$ term in the cost function.  These 26 parameters consist of 12 modulation strengths, 12 modulation phases, and 2 bias phases for the MZMs. 

As an example of the structure of the GA parameter search process, consider the first column vector in $\mathbf{C}_4$, which is $\left[ 1 \hspace{1ex} 1 \hspace{1ex}1 \hspace{1ex}1 \hspace{1ex}\right]^\text{T}$. This portion of the $\mathbf{C}_4$ matrix is responsible for generating the ideal response for the $|0\rangle$ component of the output photon's quantum state.  This column vector indicates that the amplitude of the output spectral lines for $\left[ \omega_0 \hspace{1ex} \omega_1 \hspace{1ex} \omega_2 \hspace{1ex} \omega_3 \right]^\text{T}$, which correspond to the $|0\rangle$ component of the ideal circuit response state $|\Psi_{out}\rangle$, should all be of equal amplitude and with no induced phase shift since these particular $\mathbf{C}_4$ matrix components are all real and of the same amplitude for this column vector.  Every column vector of $\mathbf{C}_4$ is non-zero and has unity amplitude, thus the ideal output response for every component of $\mathbf{C}_4$ should equally spread the amplitude of the incident photons $|i\rangle$ basis component for all $\{i|i=0,1,2,3\}$.  Induced phase shifts are either $\{0, \frac{\pi}{2}, \pi, \frac{3\pi}{2}\}$ when the $\mathbf{C}_4$ matrix component is $\{1, i, -1, -i\}$ respectively.

\begin{figure}[h!]
    \centering
    \begin{minipage}{0.45\textwidth}
        \centering
        \includegraphics[width=\linewidth]{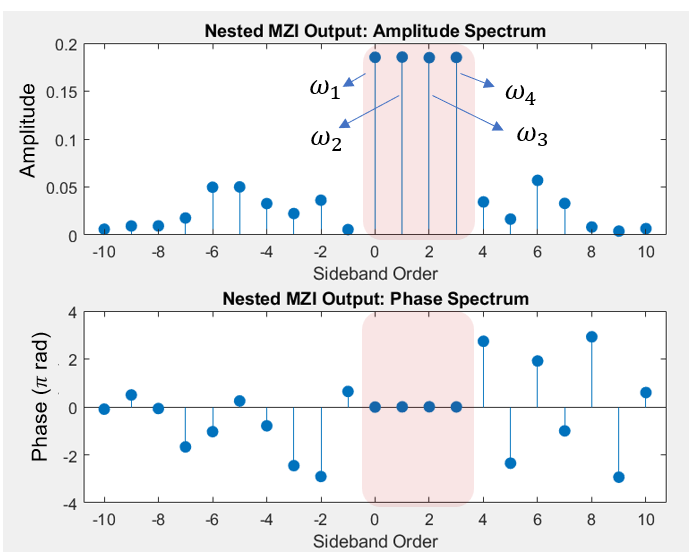}
        \captionof{figure}{Generated sidebands of the first mixer corresponding to the first column of the $\mathbf{C}_4$ gate, showing their targeted amplitudes and phases. The 26 mixer parameters are obtained using GA.}
        \label{fig:col1_C4}
    \end{minipage}
    \hfill
    \begin{minipage}{0.45\textwidth}
        \centering
        \includegraphics[width=\linewidth]{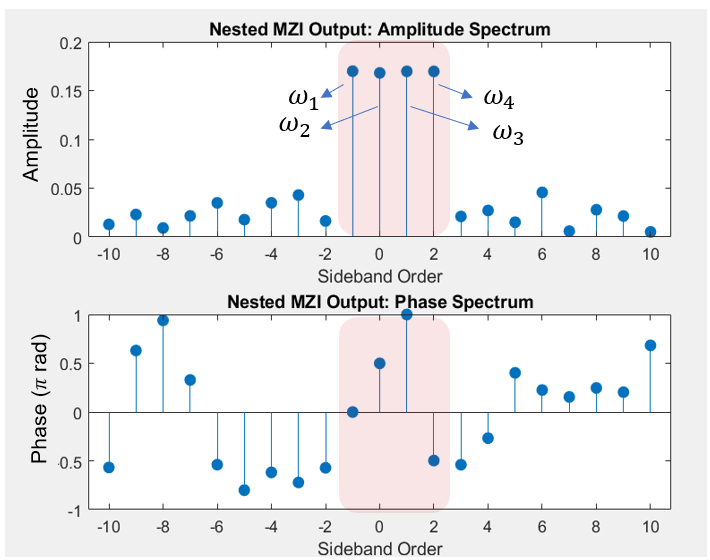}
        \captionof{figure}{Generated sidebands of the second mixer corresponding to the second column of the $\mathbf{C}_4$ gate, showing their targeted amplitudes and phases. The 26 mixer parameters are obtained using GA.}
        \label{fig:col2_C4}
    \end{minipage}
\end{figure}

We used the GA methods in MATLAB’s GA toolbox \cite{mathworks_GA}, we obtain optimized values for the 26 variables arising from the nested MZM structures in Figure \ref{fig:diagram3} that comprise the parallel branches of the QQFP shown in Figure \ref{fig:QQFP} resulting in the targeted frequency sidebands and desired amplitude and phase values shown in Figure~\ref{fig:col1_C4}.

For the second parallel nested MZM branch from the top in Figure~\ref{fig:QQFP}, that corresponds to the second column vector of the $\mathbf{C}_4$ in the QQFP gate, the order of the targeted frequency sidebands is mapped to the frequency spectrum components at indices $\left[-1, 0, 1, 2\right]$ in Figure~\ref{fig:col2_C4}. All of these spectral lines should have equal amplitude in the ideal case and their phases should shift by $\left[
0, \frac{\pi}{2}, \pi, \frac{3\pi}{2}\right]$ respectively that corresponds to the second column vector of $\mathbf{C}_4$ which is $\left[ 1 \hspace{1ex} i \hspace{1ex} -1 \hspace{1ex} -i \right]^\text{T}$. The resulting amplitude and phase of the targeted frequency bins obtained through GA optimization are shown in Figure~\ref{fig:col2_C4}.

\begin{figure}  [h!]
    \begin{minipage}{0.45\textwidth}
        \centering
        \includegraphics[width=\linewidth]{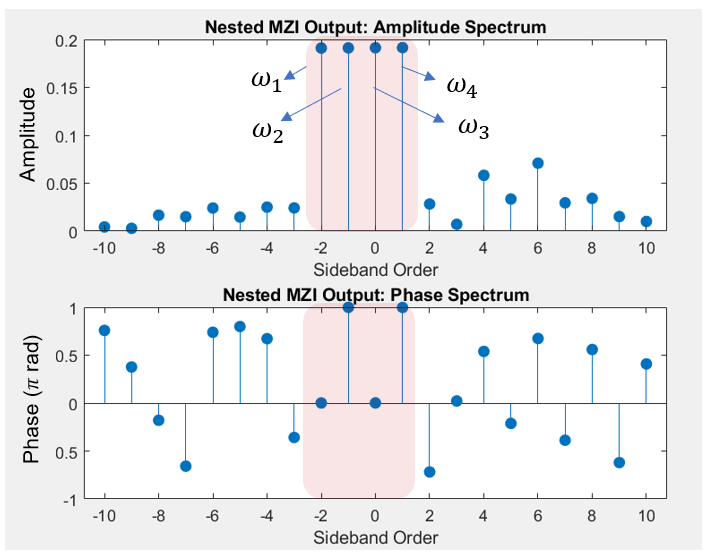}
        \captionof{figure}{Generated sidebands of the third mixer corresponding to the third column of the $\mathbf{C}_4$ gate, showing their targeted amplitudes and phases. The 26 mixer parameters are obtained using GA.}
        \label{fig:col3_C4}
    \end{minipage}
    \hfill
    \begin{minipage}{0.45\textwidth}
        \centering
        \includegraphics[width=\linewidth]{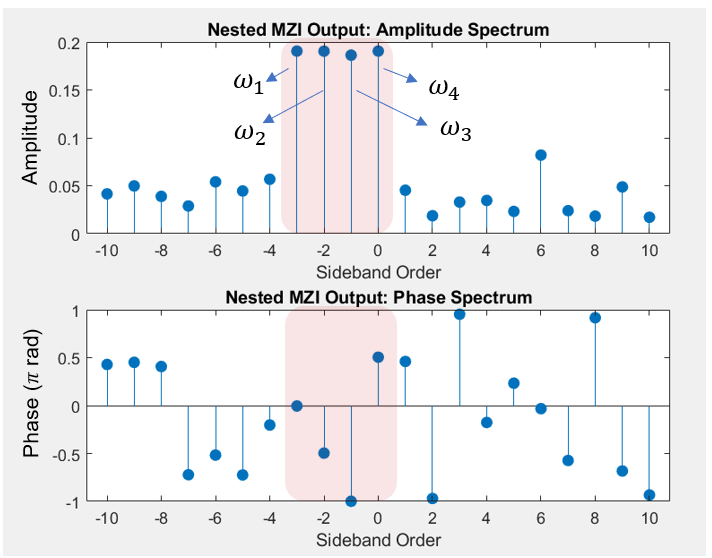}
        \captionof{figure}{Generated sidebands of the fourth mixer corresponding to the fourth column of the $\mathbf{C}_4$ gate, showing their targeted amplitudes and phases. The 26 mixer parameters are obtained using GA.}
        \label{fig:col4_C4}
    \end{minipage}
\end{figure}

Similarly, for the third parallel nested MZM branch from the top in Figure~\ref{fig:QQFP}, that corresponds to the third column vector of the $\mathbf{C}_4$ in the QQFP gate, the order of the targeted frequency sidebands is mapped to the frequency spectrum components at indices $\left[-2, -1, 0, 1\right]$ in Figure~\ref{fig:col3_C4}. All of these spectral lines should have equal amplitude in the ideal case and their phases should shift by $\left[
0, \pi, 0, \pi\right]$ respectively that corresponds to the third column vector of $\mathbf{C}_4$ which is $\left[ 1 \hspace{1ex} -1 \hspace{1ex} 1 \hspace{1ex} -1 \right]^\text{T}$. The resulting amplitude and phase of the targeted frequency bins obtained through GA optimization are shown in Figure~\ref{fig:col3_C4}.

And finally, for the fourth parallel nested MZM branch from the top in Figure~\ref{fig:QQFP}, that corresponds to the fourth column vector of the $\mathbf{C}_4$ in the QQFP gate, the order of the targeted frequency sidebands is mapped to the frequency spectrum components at indices $\left[-3, -2, -1, 0\right]$ in Figure~\ref{fig:col4_C4}. All of these spectral lines should have equal amplitude in the ideal case and their phases should shift by $\left[
0, \frac{3\pi}{2}, \pi, \frac{\pi}{2}\right]$ respectively that corresponds to the fourth column vector of $\mathbf{C}_4$ which is $\left[ 1 \hspace{1ex} -i \hspace{1ex} -1 \hspace{1ex} i \right]^\text{T}$. The resulting amplitude and phase of the targeted frequency bins obtained through GA optimization are shown in Figure~\ref{fig:col4_C4}.

\section{Analytic Approximation}
\subsection{Truncated Fourier Series}

An approximate analytic solution may be obtained by expanding the expression for the field after each phase modulator into a Fourier series. For the three tone RF modulating signal, \(m(t) \), in Eq.~\eqref{eq:fourier-series} above, the system is characterized by frequencies:
\begin{align}
\omega_1 &= \omega_i :=\text{ input optical frequency}\\
\omega_2 &= \omega_i + \delta \\
\omega_3 &= \omega_i + 2\delta \\
\omega_4 &= \omega_i + 3\delta
\end{align}

Where the output fields for the two arms are: \\

\textbf{First (Upper) Arm Output:}
\begin{equation}
\label{eq:output-field1}
E_{out,1}(t) = E_0 \sum_{n,m,k} J_n(\beta_1)J_m(\beta_2)J_k(\beta_3) \cos(\omega_1 t + n\omega_2 t + m\omega_3 t + k\omega_4 t + n\phi_1 + m\phi_2 + k\phi_3)
\end{equation}

\textbf{Second (Lower) Arm Output:}
\begin{equation}
\label{eq:output-field2}
E_{out,2}(t) = E_0 \sum_{n,m,k} J_n(\beta_4)J_m(\beta_5)J_k(\beta_6) \cos(\omega_1 t + n\omega_2 t + m\omega_3 t + k\omega_4 t + n\phi_4 + m\phi_5 + k\phi_6 + \Delta\phi)
\end{equation}

Hence, the total output field is given by summing \eqref{eq:output-field1} and \eqref{eq:output-field2}:
\begin{equation}
\label{eq:output-field_tot}
E_{out}(t)=E_{out,1}(t)+E_{out,2}(t)
\end{equation}

Expanding this into a cosine series and neglecting all but the first three terms we write, 
\begin{equation}
\label{eq:output-field_tot_fourier}
E_{out}(t) \approx A_1\cos(\omega_1t+\theta_1)+A_2\cos(\omega_2t+\theta_2)+A_3\cos(\omega_3t+\theta_3)+A_4\cos(\omega_4t+\theta_4)
\end{equation}

The Fourier coefficients are extracted from each arm separately using orthogonality. For  frequency component $h$ and arm $k$:
\begin{equation}
\label{eq:fourier_coeff}
A_{h,k} = \int_{-\infty}^{\infty} E_{out,k}(t) \cos([\omega_i + (h-1)\delta]t) \, dt
\end{equation}

For the Mach-Zehnder Modulator shown in Figure~\ref{fig:diagram2}, the four output Fourier coefficients are:

\begin{multline}
\label{eq:carrier_coeff}
\text{input color: } A_1^\text{total} = A_\text{1,1} + A_\text{1,2} = \frac{E_0}{4}\Big[J_0(\beta_1)J_0(\beta_2)J_0(\beta_3) + 2J_1(\beta_1)J_1(\beta_3)J_2(\beta_2)\cos(\phi_1 - 2\phi_2 + \phi_3) \\
+ J_0(\beta_4)J_0(\beta_5)J_0(\beta_6)\cos(\Delta\phi) + J_1(\beta_4)J_1(\beta_6)J_2(\beta_5)\cos(\Delta\phi - \phi_4 + 2\phi_5 - \phi_6) \\
+ J_1(\beta_4)J_1(\beta_6)J_2(\beta_5)\cos(\Delta\phi + \phi_4 - 2\phi_5 + \phi_6)\Big]
\end{multline}

\begin{multline}
\label{eq:1st_coeff}
\text{First Sideband: } A_2^\text{total} = A_\text{2,1} + A_\text{2,2} = \frac{E_0}{4} \Big[
-J_2(\beta_1)J_3(\beta_2)J_1(\beta_3)\cos(-2\phi_1 + 3\phi_2 - \phi_3) \\
-J_1(\beta_1)J_1(\beta_2)J_0(\beta_3)\cos(-\phi_1 + \phi_2) -
J_0(\beta_1)J_1(\beta_2)J_1(\beta_3)\cos(-\phi_2 + \phi_3) \\-
J_1(\beta_1)J_3(\beta_2)J_2(\beta_3)\cos(\phi_1 - 3\phi_2 + 2\phi_3) - J_0(\beta_6)J_1(\beta_4)J_1(\beta_5)\cos(\Delta\phi - \phi_4 + \phi_5) \\
- J_1(\beta_6)J_2(\beta_4)J_3(\beta_5)\cos(\Delta\phi - 2\phi_4 + 3\phi_5 - \phi_6) - J_0(\beta_4)J_1(\beta_5)J_1(\beta_6)\cos(\Delta\phi - \phi_5 + \phi_6) \\
- J_1(\beta_4)J_2(\beta_6)J_3(\beta_5)\cos(\Delta\phi + \phi_4 - 3\phi_5 + 2\phi_6)\Big]
\end{multline}

\begin{multline}
\label{eq:2nd_coeff}
\text{Second Sideband: } A_3^\text{total} = A_\text{3,1} + A_\text{3,2} = \frac{E_0}{4}\Big[
J_4(\beta_1)J_6(\beta_2)J_2(\beta_3)\cos(-4\phi_1 + 6\phi_2 - 2\phi_3) +\\
J_2(\beta_1)J_2(\beta_2)J_0(\beta_3)\cos(-2\phi_1 + 2\phi_2) +
J_0(\beta_1)J_2(\beta_2)J_2(\beta_3)\cos(-2\phi_2 + 2\phi_3)\\
+ J_0(\beta_6)J_2(\beta_4)J_2(\beta_5)\cos(\Delta\phi - 2\phi_4 + 2\phi_5) - J_0(\beta_5)J_1(\beta_4)J_1(\beta_6)\cos(\Delta\phi - \phi_4 + \phi_6) \\
+ J_0(\beta_4)J_2(\beta_5)J_2(\beta_6)\cos(\Delta\phi - 2\phi_5 + 2\phi_6)\Big]
\end{multline}

\begin{multline}
\label{eq:3rd_coeff}
\text{Third Sideband: } A_4^\text{total} = A_\text{4,1} + A_\text{4,2} = \frac{E_0}{4}\Big[
-J_4(\beta_1)J_5(\beta_2)J_1(\beta_3)\cos(-4\phi_1 + 5\phi_2 - \phi_3) \\-
J_3(\beta_1)J_3(\beta_2)J_0(\beta_3)\cos(-3\phi_1 + 3\phi_2) +
J_2(\beta_1)J_1(\beta_2)J_1(\beta_3)\cos(-2\phi_1 + \phi_2 + \phi_3) \\+
J_1(\beta_1)J_1(\beta_2)J_2(\beta_3)\cos(-\phi_1 - \phi_2 + 2\phi_3) - J_0(\beta_6)J_3(\beta_4)J_3(\beta_5)\cos(\Delta\phi - 3\phi_4 + 3\phi_5) \\
+ J_1(\beta_5)J_1(\beta_6)J_2(\beta_4)\cos(\Delta\phi - 2\phi_4 + \phi_5 + \phi_6) + J_1(\beta_4)J_1(\beta_5)J_2(\beta_6)\cos(\Delta\phi - \phi_4 - \phi_5 + 2\phi_6) \\
- J_0(\beta_4)J_3(\beta_5)J_3(\beta_6)\cos(\Delta\phi - 3\phi_5 + 3\phi_6)\Big]
\end{multline}

\subsection{Sensitivity Analysis}

A key advantage of deriving this approximate analytic solution is its utility in assessing the sensitivity of the results to perturbations in the governing parameters. In this section, we present representative outcomes focused on one of the newly generated frequency components, as described in Eq.~\eqref{eq:1st_coeff} above. We formulate the sensitivity expressions and illustrate them graphically across an expanded range of the relevant variables. The specific parameters analyzed here encompass variations in the relative phase, $\Delta\phi$, between the upper and lower arms of the MZM, fluctuations in the amplitude, $\beta_5$, of the modulation signal $m(t)$, and shifts in the RF signal phases, $\phi_1$ and $\phi_2$.

In general, we employ a normalized Jacobian, such that the phase sensitivity is defined as:

\begin{equation}
\label{eq:S_phi}
S_\phi = \frac{1}{A_{1st}} \frac{\partial A_{1st}}{\partial \phi}
\end{equation}

and the RF amplitude sensitivity is expressed as:
\begin{equation}
\label{S_beta}
S_\beta = \frac{\beta}{A_{1st}} \frac{\partial A_{1st}}{\partial \beta}
\end{equation}

We begin with the relative phase, $\Delta\phi$, between the upper and lower arms of the MZM. The explicit form for this sensitivity, $S_{\Delta\phi}$, is
\begin{align}
\label{eq:S_delta_phi}
S_{\Delta\phi} = \frac{
\begin{aligned}
&-[J_1(\beta_6) [ J_2(\beta_4) J_3(\beta_5) \sin(\Delta\phi - 2\phi_4 + 3\phi_5 - \phi_6) + J_0(\beta_4) J_1(\beta_5) \sin(\Delta\phi - \phi_5 + \phi_6)] \\
&+ J_1(\beta_4) J_3(\beta_5) J_2(\beta_6) \sin(\Delta\phi + \phi_4 - 3\phi_5 + 2\phi_6) 
+ J_1(\beta_4) J_1(\beta_5) J_0(\beta_6) \sin(\Delta\phi - \phi_4 + \phi_5)]
\end{aligned}
}{
\begin{aligned}
&J_2(\beta_1) J_3(\beta_2) J_1(\beta_3) \cos(2\phi_1 - 3\phi_2 + \phi_3)
+ J_1(\beta_1) J_3(\beta_2) J_2(\beta_3) \cos(\phi_1 - 3\phi_2 + 2\phi_3) \\
&+ J_1(\beta_1) J_1(\beta_2) J_0(\beta_3) \cos(\phi_1 - \phi_2)
+ J_0(\beta_1) J_1(\beta_2) J_1(\beta_3) \cos(\phi_2 - \phi_3) \\
&+ J_2(\beta_4) J_3(\beta_5) J_1(\beta_6) \cos(\Delta\phi - 2\phi_4 + 3\phi_5 - \phi_6)
+ J_1(\beta_4) J_3(\beta_5) J_2(\beta_6) \cos(\Delta\phi + \phi_4 - 3\phi_5 + 2\phi_6) \\
&+ J_1(\beta_4) J_1(\beta_5) J_0(\beta_6) \cos(\Delta\phi - \phi_4 + \phi_5) 
+J_0(\beta_4) J_1(\beta_5) J_1(\beta_6) \cos(\Delta\phi - \phi_5 + \phi6)
\end{aligned}
}
\end{align}

\begin{figure}[h!]
    \centering
    \begin{minipage}{0.45\textwidth}
        \centering
        \includegraphics[width=\linewidth]{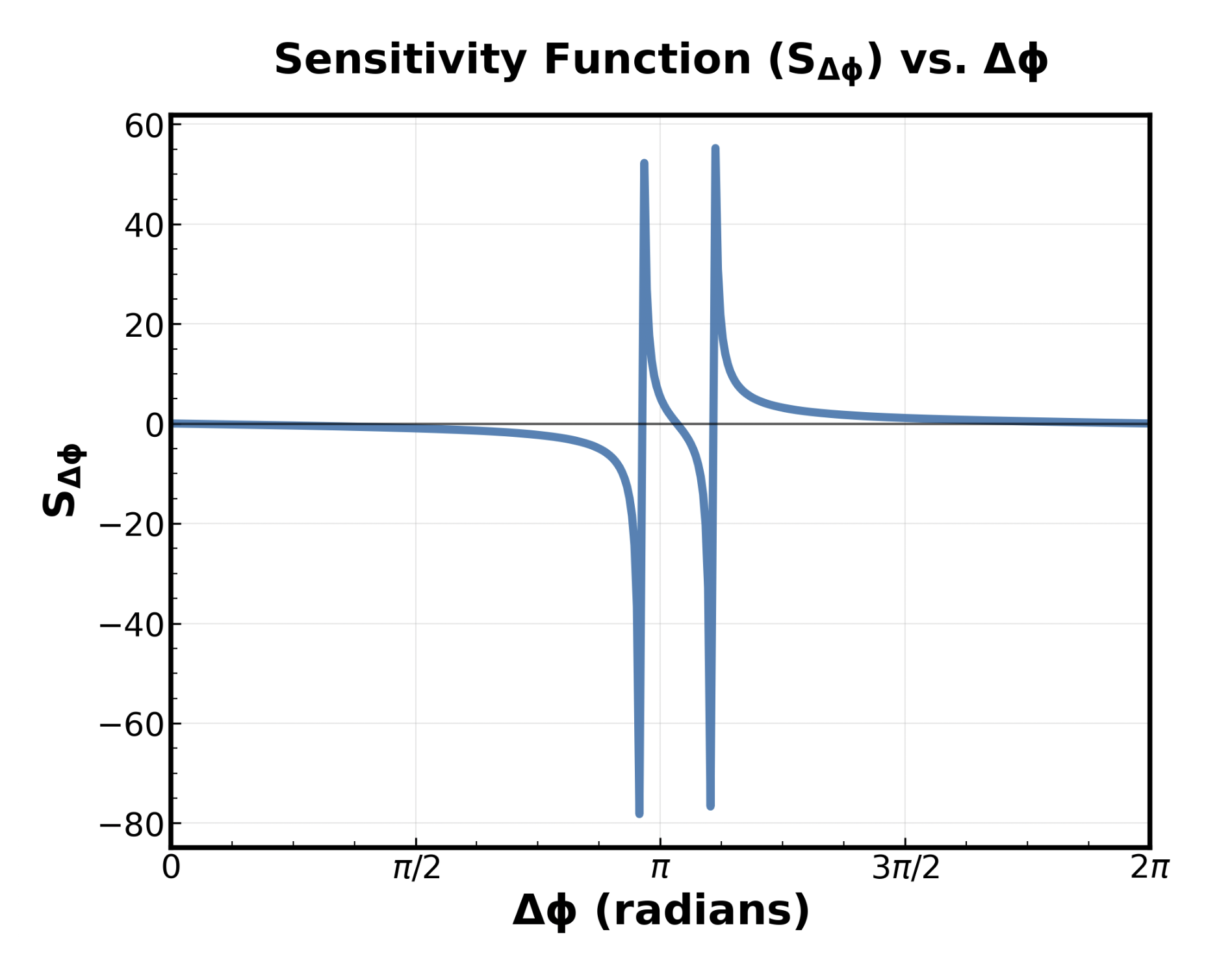}
        \captionof{figure}{Plot of sensitivity function: $S_{\Delta\phi}$ vs. $\Delta\phi$.}
        \label{fig:S_Dphi}
    \end{minipage}
    \hfill
    \begin{minipage}{0.45\textwidth}
        \centering
        \includegraphics[width=\linewidth]{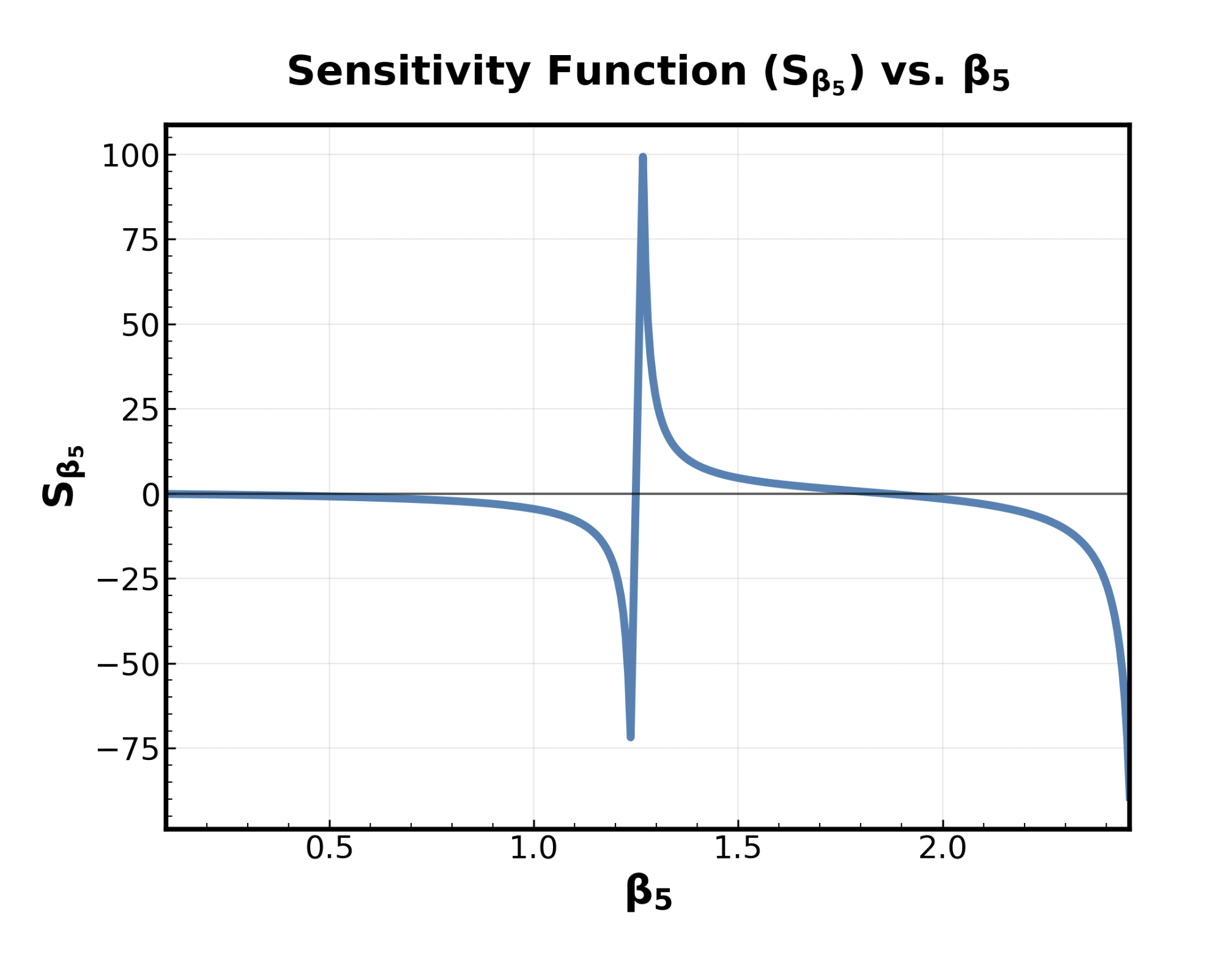}
        \captionof{figure}{Plot of sensitivity function: $S_{\beta_5}$ vs. $\beta_5$.}
        \label{fig:S_beta5}
    \end{minipage}
\end{figure}

This is depicted in Figure~\ref{fig:S_Dphi}. Among the three categories of sensitivity parameters, this relative phase is arguably the most straightforward to manage, as it involves a single DC bias voltage applied to one arm of the MZM or a differential DC bias across both arms. In either scenario, a standard 12-bit Digital-to-Analog Converter (DAC) should suffice.

The next parameter in terms of control feasibility is the amplitude of the applied RF signal. The sensitivity expression for the first sideband with respect to $\beta_5$ is:
\begin{align}
\label{eq:S_beta_5}
    S_{\beta_5} = \frac{
\begin{aligned}
&\beta_5[J_2(\beta_4)[J_2(\beta_5) - J_4(\beta_5)] J_1(\beta_6) \cos(\Delta\phi - 2\phi_4 + 3\phi_5 - \phi_6)\\ &+ J_1(\beta_4)[J_2(\beta_5) - J_4(\beta_5)] J_2(\beta_6) \cos(\Delta\phi + \phi_4 - 3\phi_5 + 2\phi_6)\\
&+ J_1(\beta_4)[J_0(\beta_5) - J_2(\beta_5)] J_0(\beta_6) \cos(\Delta\phi - \phi_4 + \phi_5) + J_0(\beta_4)[J_0(\beta_5) - J_2(\beta_5)] J_1(\beta_6) \cos(\Delta\phi - \phi_5 + \phi_6)]
\end{aligned}
}{
\begin{aligned}
&2[J_2(\beta_1) J_3(\beta_2) J_1(\beta_3) \cos(2\phi_1 - 3\phi_2 + \phi_3)
+ J_1(\beta_1) J_3(\beta_2) J_2(\beta_3) \cos(\phi_1 - 3\phi_2 + 2\phi_3) \\
&+ J_1(\beta_1) J_1(\beta_2) J_0(\beta_3) \cos(\phi_1 - \phi_2)
+ J_0(\beta_1) J_1(\beta_2) J_1(\beta_3) \cos(\phi_2 - \phi_3) \\
&+ J_2(\beta_4) J_3(\beta_5) J_1(\beta_6) \cos(\Delta\phi - 2\phi_4 + 3\phi_5 - \phi_6)
+ J_1(\beta_4) J_3(\beta_5) J_2(\beta_6) \cos(\Delta\phi + \phi_4 - 3\phi_5 + 2\phi_6) \\
&+ J_1(\beta_4) J_1(\beta_5) J_0(\beta_6) \cos(\Delta\phi - \phi_4 + \phi_5) 
+J_0(\beta_4) J_1(\beta_5) J_1(\beta_6) \cos(\Delta\phi - \phi_5 + \phi_6)]
\end{aligned}
}
\end{align}
This is illustrated in Figure~\ref{fig:S_beta5}, revealing preferred operational regimes compared to others. Once again, the modulation amplitude represents a largely linear, DC-controlled quantity, making it relatively straightforward to regulate.

\begin{figure}[h!]  
    \begin{minipage}{0.45\textwidth}
        \centering
        \includegraphics[width=\linewidth]{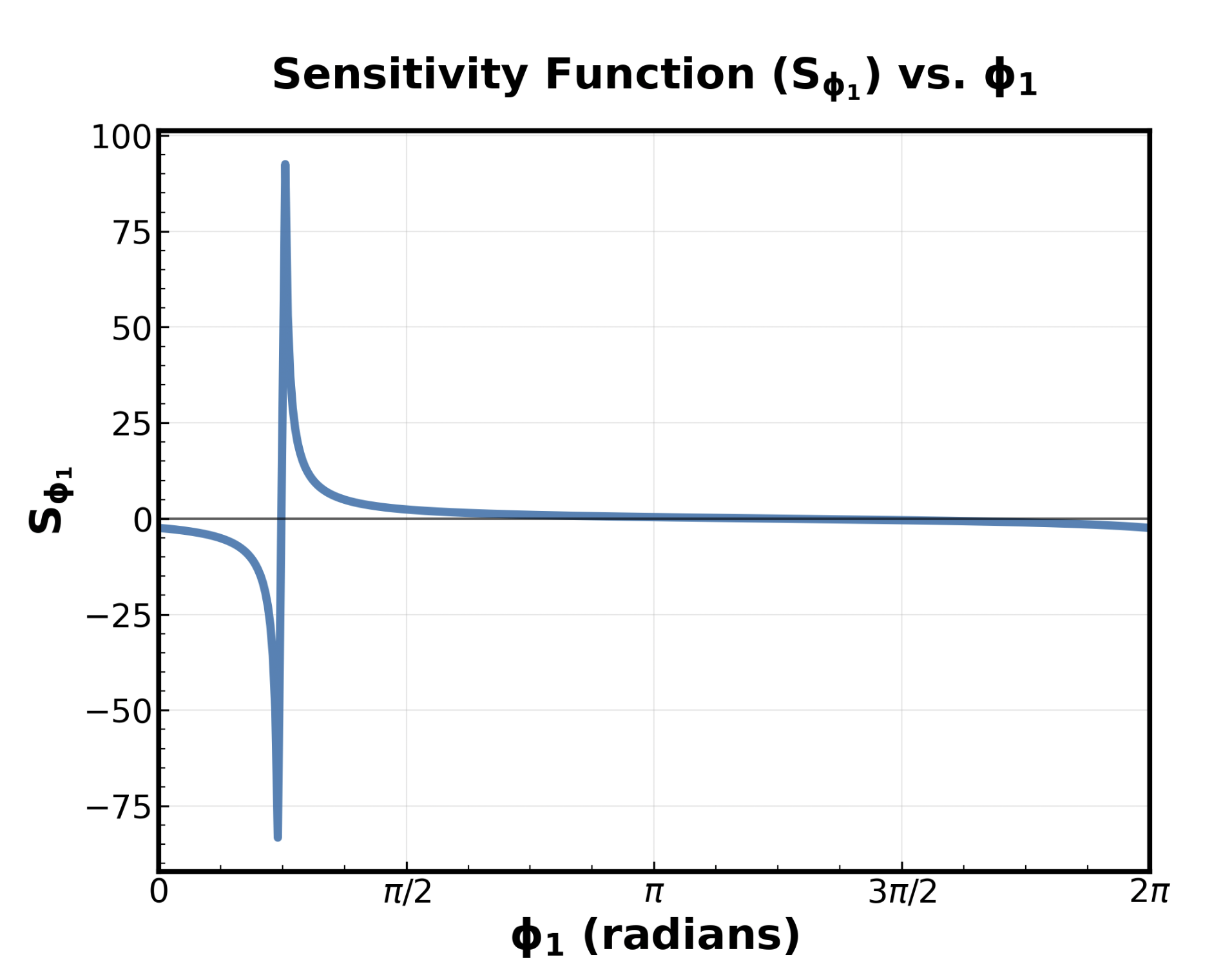}
        \captionof{figure}{Plot of sensitivity function: $S_{\phi_1}$ vs. $\phi_1$.}
        \label{fig:S_phi1}
    \end{minipage}
    \hfill
    \begin{minipage}{0.45\textwidth}
        \centering
        \includegraphics[width=\linewidth]{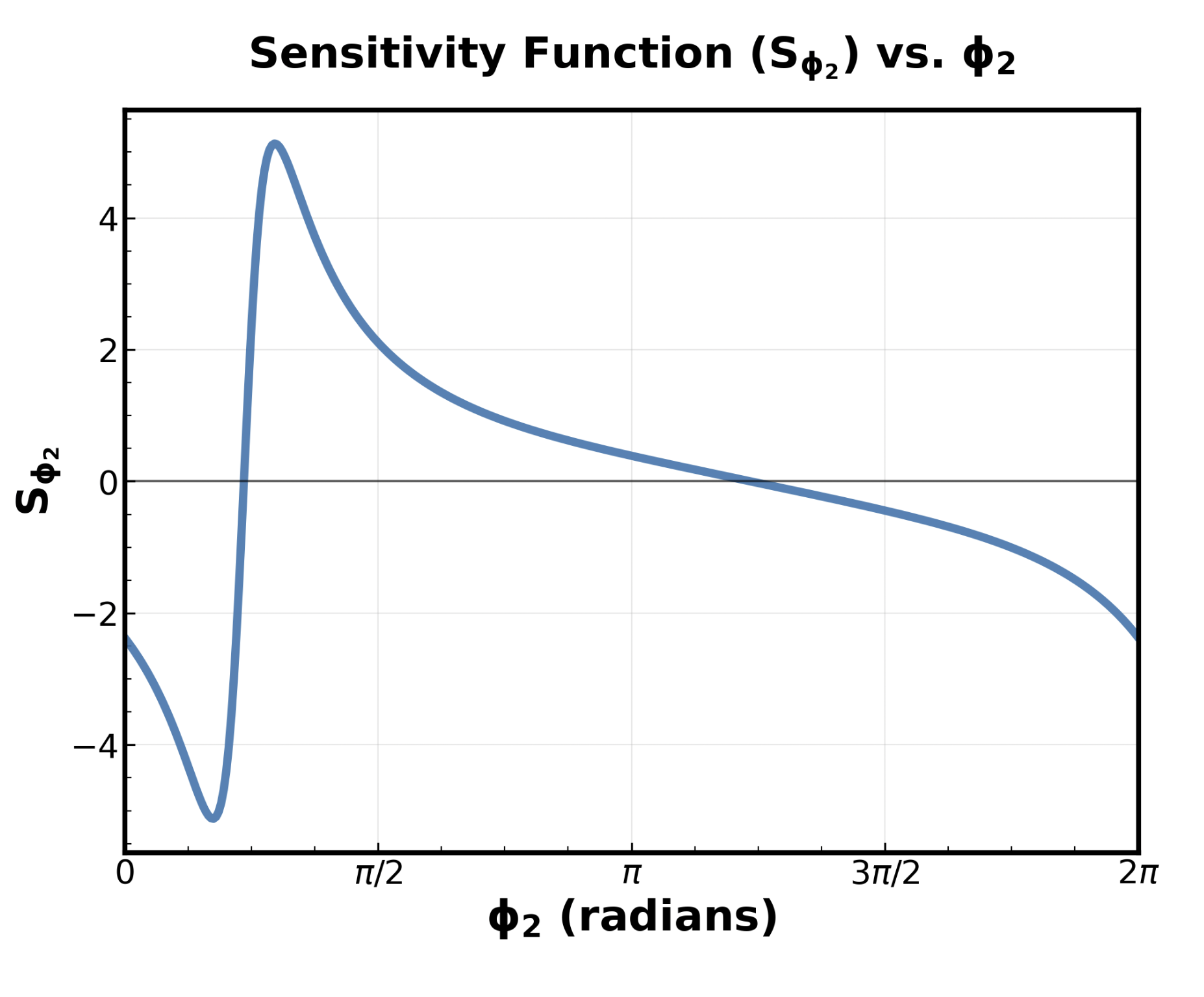}
        \captionof{figure}{Plot of sensitivity function: $S_{\phi_2}$ vs. $\phi_2$.}
        \label{fig:S_phi2}
    \end{minipage}
\end{figure}

Phase stability (or equivalently, frequency stability) of the RF source is arguably the most critical specification for the modulation system. We evaluate the sensitivity of the first sideband to both $\phi_1$ and $\phi_2$:
\begin{align}
    S_{\phi_1} = \frac{
    \begin{aligned}
        -[J_3(\beta_2)[2J_2(\beta_1)J_1(\beta_3)\sin(2\phi_1 - 3\phi_2 + \phi_3) + J_1(\beta_1)J_2(\beta_3)\sin(\phi_1 - 3\phi_2 + 2\phi_3)] \\+ J_1(\beta_1)J_1(\beta_2)J_0(\beta_3)\sin(\phi_1 - \phi_2)]
    \end{aligned}
    }{
    \begin{aligned}
        &J_2(\beta_1) J_3(\beta_2) J_1(\beta_3) \cos(2\phi_1 - 3\phi_2 + \phi_3)
+ J_1(\beta_1) J_3(\beta_2) J_2(\beta_3) \cos(\phi_1 - 3\phi_2 + 2\phi_3) \\
&+ J_1(\beta_1) J_1(\beta_2) J_0(\beta_3) \cos(\phi_1 - \phi_2)
+ J_0(\beta_1) J_1(\beta_2) J_1(\beta_3) \cos(\phi_2 - \phi_3) \\
&+ J_2(\beta_4) J_3(\beta_5) J_1(\beta_6) \cos(\Delta\phi - 2\phi_4 + 3\phi_5 - \phi_6)
+ J_1(\beta_4) J_3(\beta_5) J_2(\beta_6) \cos(\Delta\phi + \phi_4 - 3\phi_5 + 2\phi_6) \\
&+ J_1(\beta_4) J_1(\beta_5) J_0(\beta_6) \cos(\Delta\phi - \phi_4 + \phi_5) 
+J_0(\beta_4) J_1(\beta_5) J_1(\beta_6) \cos(\Delta\phi - \phi_5 + \phi6)
    \end{aligned}
    }
\end{align}

\begin{align}
    S_{\phi_2} = \frac{
    \begin{aligned}
        3J_3(\beta_2)[J_2(\beta_1)J_1(\beta_3)\sin(2\phi_1 - 3\phi_2 + \phi_3) + J_1(\beta_1)J_2(\beta_3)\sin(\phi_1 - 3\phi_2 + 2\phi_3)] \\+ J_1(\beta_1)J_1(\beta_2)J_0(\beta_3)\sin(\phi_1 - \phi_2) - 
        J_0(\beta_1)J_1(\beta_2)J_1(\beta_3)\sin(\phi_2 - \phi_3)
    \end{aligned}
    }{
    \begin{aligned}
        &J_2(\beta_1) J_3(\beta_2) J_1(\beta_3) \cos(2\phi_1 - 3\phi_2 + \phi_3)
+ J_1(\beta_1) J_3(\beta_2) J_2(\beta_3) \cos(\phi_1 - 3\phi_2 + 2\phi_3) \\
&+ J_1(\beta_1) J_1(\beta_2) J_0(\beta_3) \cos(\phi_1 - \phi_2)
+ J_0(\beta_1) J_1(\beta_2) J_1(\beta_3) \cos(\phi_2 - \phi_3) \\
&+ J_2(\beta_4) J_3(\beta_5) J_1(\beta_6) \cos(\Delta\phi - 2\phi_4 + 3\phi_5 - \phi_6)
+ J_1(\beta_4) J_3(\beta_5) J_2(\beta_6) \cos(\Delta\phi + \phi_4 - 3\phi_5 + 2\phi_6) \\
&+ J_1(\beta_4) J_1(\beta_5) J_0(\beta_6) \cos(\Delta\phi - \phi_4 + \phi_5) 
+J_0(\beta_4) J_1(\beta_5) J_1(\beta_6) \cos(\Delta\phi - \phi_5 + \phi6)
    \end{aligned}
    }
\end{align}

These sensitivities are plotted in Figures~\ref{fig:S_phi1} and \ref{fig:S_phi2}, respectively.

\section{Conclusions}

Multicolor photons are an attractive proposition because of their potential to support high-dimensional states. The parallel design facilitates scalability to similarly higher dimensions, as additional mixers can be incorporated without fundamental redesign, enabling processing of qudits with arbitrary dimensionality limited only by practical integration constraints. The scaling of a Hadamard (beamsplitter) for two-dimensional qubits to a Chrestenson (\(\bf{C}_4\)) gate serves as an illustration of this progression to higher dimension. More modulators in parallel may be added as the qudit dimension increases.

In this paper, we have introduced a new arrangement of mixers and filters to process multicolor quantum states. If previous quantum processors are viewed as a serial cascade of mixers and filters \citep{harris2008, sensarn2009, lukens2017, kues2017, lu2018, imany2018, lu2019, kues2019, lu2023}, the arrangement here may be considered to be parallel. In previous realizations, the modulator and filtering is separated into distinct stages. Here, the two functions are combined. The mixers here are placed, if you will, between the demux/mux gratings of the programmable filters in references \citep{lukens2017, kues2017, lu2018, imany2018, lu2019, kues2019, lu2023}. 

Further, in this work, the single arm phase modulators are replaced with MZMs. Again, filtering is combined with mixing, and additional degrees of freedom are included in the processor. 

The parallel configuration offers an advantage with respect to parasitic (coupling) losses. In a serial architecture, parasitic losses cascade multiplicatively. A parallel arrangement, however, distributes one parasitic loss across each component so that, after the mux, only the single parasitic loss term is present. 

The architecture introduced here lends itself as well to photonic integration. Sources, demultiplexers, modulators and multiplexers may be realized in a TFLN material system. This architecture extends naturally to other gates beyond the \(\bf{C}_4\), such as $\mathbf{S}$, $\mathbf{X}$, $\mathbf{Z}$, and $\mathbf{T}$ gates, or custom unitary operations by adjusting the RF modulation parameters and mixer configurations. Using the same generic cell in an integrated fabric suggests a quantum field programmable photonic gate array (q-FPPGA).  

In summary of results, this paper has introduced a novel parallel architecture for realizing quantum gates in frequency-comb qudit systems, demonstrated through the implementation of the Chrestenson (\(\bf{C}_4\)) gate using both numerical optimization via genetic algorithms and an analytic approximation based on truncated Fourier series. The approach leverages nested Mach-Zehnder modulators to achieve the required mixing of frequency components, with sensitivity analysis highlighting robust operational regimes against parameter variations.

\end{document}